\documentstyle[12pt]{article}

\begin{document}

\centerline{\Large Basis states for gravitons in non-perturbative loop}
\centerline{\Large representation space}
\vspace{0.5cm}
\centerline{\rm Junichi Iwasaki}
\vspace{0.2cm}
\centerline{\rm Department of Gravitation and Field Theory}
\centerline{\rm Instituto de Ciencias Nucleares}
\centerline{\rm Universidad Nacional Autonoma de Mexico}
\centerline{\rm AP 70-543, Mexico DF 04510, Mexico}
\vspace{0.2cm}
\centerline{\rm Department of Physics}
\centerline{\rm Universidad Autonoma Metropolitana, Iztapalapa}
\centerline{\rm AP 55-534, Mexico DF 09340, Mexico}
\vspace{0.2cm}
\centerline{\rm E-mail:\ iwasaki@abaco.uam.mx}
\vspace{0.2cm}
\centerline{June 8, 1998}


\begin{abstract}
A relation between the non-perturbative loop representation space 
(for use of exact quantization of general relativity) 
and the semi-classical loop representation space 
(for use of approximate quantization of general relativity 
with a flat background spacetime) 
is studied.
A sector of (approximate) states and a sector of operators in the
non-perturbative loop representation space are made related respectively
to the physical states and the basic variables of the semi-classical
loop representation space through a transformation.
This transformation makes a construction of graviton states within the 
non-perturbative theory possible although the notion of gravitons
originally emerged from semi-classical perturbative treatments.
This transformation is ``exact'' in the sense that it does not contain
an error term, a fact contrast to the previous construction of 
a similar transformation.
This fact allows the interpretation that these graviton states represent
free gravitons even after introducing the self-interaction of
gravitons into the theory; the presence of an error term
of order of possible perturbative self-interaction would spoil
this interpretation.
The existence of such an ``exact'' relation of the two theories supports
the potential ability of the non-perturbative loop representation 
quantum gravity
to address the physics of gravitons, namely quanta for large scale small
fluctuations of gravitational field from the flat background spacetime.  
\end{abstract}


\section{Introduction}\label{sec:introduction}

In order to quantize general relativity \cite{ashtekar}, 
the so-called loop representation \cite{gambini,loop} has been applied 
both non-perturbatively (exactly) \cite{loop}
and semi-classically (approximately) \cite{linear}.
The latter is an approximation of the former and the physics described by
the latter must be contained in the former if they are consistently
related to each other.
However, in order to quantize semi-classically, one first linearizes 
general relativity at the classical level and then quantizes it. 
The resulting theory is 
mathematically different from the non-perturbatively quantized theory
of exact general relativity.
To find the mathematical structure of the semi-classical theory within
the non-perturbative theory is highly non-trivial task.
This seems a necessary step in order to say:
this non-perturbative theory is really an ``exact'' physics theory and
this semi-classical theory is really an ``approximate'' physics theory.
We study a possible way of relating the two theories.

This study was initiated a few years ago \cite{mapM,vacuum}.
(An outline of the work is mentioned in section \ref{sec:comparison}.)
However, it was thought that the relation between the two theories was just 
an approximation useful for ``estimating'' the approximate free graviton 
states and that the inclusion of self-interaction spoiled the approximation.
In order to include the self-interaction effects, one had to wait for the 
development of computation techniques for non-perturbative
transition amplitudes such as regularizations of the exact Hamiltonian
constraint \cite{thiemann,algebra,consistency} 
and sum-over-surfaces formulations 
\cite{worldsheet,surface,causal,relative,foam,iwasaki}.

Recently \cite{jalapa}, 
it was re-examined and speculated that there might exist 
an ``exact'' relation between the two theories when considered only at
the vicinity of the flat spacetime.
(The ``exact'' does not mean the two theories have the same 
mathematical structure.
This ``exactness'' should be understood in a similar sense that any
spacetime metric $g_{\mu\nu}$ can be written as the sum of the flat
spacetime metric and a deviation from it, namely
$g_{\mu\nu}=\eta_{\mu\nu}+h_{\mu\nu}$.
This can be done always and exactly and this fact itself does not mean
that this splitting is an approximation.
However, this splitting of metric is physically sensible as 
an approximation only if $h_{\mu\nu}$ is small enough compared to unity.)
If it is true, then not only can the free graviton states determined 
in the non-perturbative theory be kept meaningful after the inclusion of 
self-interaction but also may the relation between the two theories
allow to include the self-interaction effects ``perturbatively''
within the non-perturbative representation space \cite{loop-gravitons}.

We study in this paper such an ``exact'' relation of the two theories.
We present a closed form of the relation of states of the two theories
without error terms.
It was previously thought that error terms were inavoidable and 
would make the inclusion
of self-interaction terms in the Hamiltonian obscure.
Although the relation has a closed form, still it is an approximation
meaningful at the vicinity of the flat spacetime.
Possible applications of the relation
such as computations of graviton-graviton scattering amplitudes would be 
understood as approximation schemes as a tentative step toward the 
developments of non-perturbative computation techniques 
(see \cite{review} for recent review in this direction).

Some of the important consequences of the relation between the 
non-perturbative and semi-classical theories are the following.
The semi-classical theory has a background flat spacetime (a classical 
solution) but the non-perturbative theory does not.
The relation we study requires the presence of a particular spin-network,
called ``weave'' \cite{weave,norbert},
in the non-perturbative representation space.
The weave presents a discrete Planck scale structure but approximates 
the flat space when probed at scales far larger than the Planck length.
In a sense the weave represents a classical limit of the theory.
Because of the presence of the weave, we can show the existence of 
an explicit relation of the two theories meaning that the non-pertubative 
theory contains the graviton physics described by 
the semi-classical theory.
In other words, the existence of the relation of the two theories supports
the existence of a possible discrete Planck scale structure of space
presented by the weave 
without conflictions with large scale continuum pictures.
The Planck scale corrections due to the discrete structure of the weave 
is absorbed to the Planck length constant to redefine 
a ``coarse-grained'' constant and hence the Planck scale structure
of the weave cannot be seen from the semi-classical theory.

The graviton states determined within the non-perturbative theory are
quantum states dressed with virtual loop gravitons around the weave
and can be seen as quantum states of the semi-classical theory dressed
with virtual gravitons around the flat space when probed at large scales.
They are supposed to contain dynamical information, although it is an 
approximation sensible at the vicinity of flat spacetime, 
and hence they toghther with possible
graviton-graviton scattering amplitudes between them represent the flat 
spacetime with large scale small fluctuations of gravitational field, 
namely gravitons.

This is a physical picture emerging from the non-perturbative loop
representation space in addition to other physical aspects 
\cite{discrete,loll,geometry,area,black-hole,entropy}.

\subsection*{Illustration of the idea}

In the rest of the introduction section
we illustrate the idea we follow in terms of a simple model; namely 
quantum mechanics of a non-relativistic particle in one-dimensional space.
It is then followed by an outline of the following sections, where
we make our discussions parallel to the illustration.

The Hamiltonian of a particle with a mass $m$ at position $X$ in a potential 
$V(X)$ is given by
$H:={P^2\over 2m}+V(X)$.
Here, $P$ is the momentum of the particle and is an operator satisfying
the commutation relation $[X,P]=i\hbar$ in a representation space.
We work in a representation in which $X$ is a multiplication number and
states are function of $X$.

In general, to find eigen states of the Hamiltonian is difficult and one
needs an approximation in order to discuss physics.
Let us suppose that the potential has a minimum at the origin $X=0$ 
and varies slowly at the vicinity of the origin. 
We specify the vicinity by the range
$-\varepsilon<X<\varepsilon$ with a positive small real number 
$\varepsilon$.
Rewrite the potential as an expansion about the origin
$V(X)=V(0)+{1\over2}V''(0)X^2+{\cal O}(\varepsilon^3)$,
and define a Hamiltonian approximating the exact Hamiltonian as
$H':={p^2\over 2m}+{1\over2}m\omega^2x^2$.
Here, $\omega$ is a constant defined such that $m\omega^2=V''(0)$ and
$x$ is the position of the particle relative to the position of
the minimum of the potential, namely $x:=X$.
($x=X$ exactly holds always, but the approximate model is physically 
sensible only when $x$ is in the range.)
The momentum $p$ is an operator satisfying the commutation relation
$[x,p]=i\hbar$ in a representation space.
We work in a representation in which $x$ is a multiplication number and
states are functions of $x$.
In general, the two representation spaces are different and 
we have to know their relation in order to discuss physics approximating 
some aspect of the exact model.

Given a state of the exact model $f(X)$, define a transformation
$\exp[{i\over\hbar}x\cdot P]$.
When applied to the state $f(X)$, this transformation produces a state
$f(x)$ of the approximate model if evaluated at the origin $X=0$.
This is the relation of the states in the two representation spaces.
The function of the corresponding state of the approximate model is 
the same as the function of the exact model with $X$ replaced by $x$.
But we know $x=X$ by definition and hence the corresponding states 
in the two models are exactly identical.
This seems a trivial result.
But if we go to quantum gravity, which is a field theory and
has an infinite number of degrees of freedom, 
then an analogous consideration gives non-trivial consequences.

Since the approximate model can be solved exactly, a solution $\psi(x)$
can be understood as one corresponding to a state $\psi(X)$ of the exact
model and the latter can be interpreted as an approximate solution 
of the exact model.
Its physical meaning is the same as the physical meaning of $\psi(x)$
of the approximate model and this physics can be understood as contained 
in the exact model.

In the following sections we discuss a relation between the 
non-perturbative and semi-classical theories of loop quantum gravity.
(The relation is an analogue of the relation between the two models
in the illustration.)
In section \ref{sec:non-perturbative} we define basis states and
functions of them in the non-perturbative theory
(the analogues of $X$ and $f(X)$ of the exact model in the illustration).
In section \ref{sec:semi-classical} we review the basis states and
functions of them of the semi-classical theory
(the analogues of $x$ and $f(x)$ of the approximate model in 
the illustration).
In section \ref{sec:transformation} we define a transformation from 
one basis to the other of the two theories
(the analogue of $\exp[{i\over\hbar}x\cdot P]$ in the illustration).
Then show that the analytic function of the basis states in the 
non-perturbative theory is transformed to the same analytic function of
the basis states of the semi-classical theory if evaluated at the weave
corresponding to flat space.
(This is analogous to the fact that $f(x)$ is transformed to $f(X)$
through $\exp[{i\over\hbar}x\cdot P]$ if evaluated 
at the origin $X=0$ in the illustration.)
The transformation  can be defined at any spin-network ``exactly''
without error terms but it is physically sensible as an approximation
when it is evaluated at the weave.
(This situation is analogous to the fact that $\exp[{i\over\hbar}x\cdot P]$
can be defined at any position $X$ exactly without error terms but it is
physically sensible as an approximation when it is evaluated at the origin
$X=0$.)
We compare the present work with the previous work in section 
\ref{sec:comparison}.
We conclude our study at the end.


\section{Basis states in the non-perturbative theory}
\label{sec:non-perturbative}

The states in the non-perturbative theory are functions on spin-networks
$\Gamma$, denoted by $\Psi(\Gamma)=\langle\Gamma|\Psi\rangle$ 
\cite{spin-network}.

We restrict ourselves to a family of states, which serve as our domain 
states for gravitons, as follows.
\begin{eqnarray}
&&
\langle\Gamma|f(G^\pm)\rangle:=
N^{-1}(\Gamma)\sum_{\{\gamma\}\in\Gamma}
(-2)^{n(\{\gamma\})-1}(-1)^{c(\{\gamma\})} f(G^\pm[\{\gamma\}]).
\end{eqnarray}
Here $\{\gamma\}$ is a set of single loops made out of all the segments
belonging to the edges of $\Gamma$ and the $n(\{\gamma\})$ is the
number of single loops in $\{\gamma\}$.
{}From a given spin-network $\Gamma$, a finite number of sets of
single loops can be constructed as follows.
Place $2j$ identical segments (or lines) on an edge with spin-$j$.
Repeat for the other edges of $\Gamma$.
By connecting every two segments (not on the same edge) meeting 
at a vertex, fix a set of single loops out of all the segments.
Permutate the connectivity of
two segments on one edge of $\Gamma$.
In other words, choose two segments on a single edge and cut them
at the middle, then reconnect them in the alternative way so that
any of the resulting segments does not retrace its way.
This permutation creates another set of single loops.
Repeat permutations of the connectivity
of the segments on the edge in all the possible ways.
The number of the permutations of segments performed on 
a single edge with spin-$j$ is $(2j)!$.
Do the same thing for all the edges.
This procedure creates all the possible sets of single loops out of
the segments belonging to the edges of $\Gamma$.
$\{\gamma\}\in\Gamma$ means one of the sets defined from $\Gamma$
and $c(\{\gamma\})$ is
the number of the permutations performed to define $\{\gamma\}$.
$N(\Gamma)$ is the number of all the possible permutations for $\Gamma$,
$N(\Gamma):=\prod_{e\in\Gamma}(2j_e)!$;
$e\in\Gamma$ means an edge $e$ in $\Gamma$ and $2j_e$ is the number
of the segments placed on the edge $e$ 
($j_e$ is the spin of the edge $e$).
If $\Gamma$ has a vertex with valence more than three, we understand that
the vertex is made of a trivalent spin-network with ``virtual'' edges
\cite{spin-network} representing an intertwiner consistent with the 
adjacent edges to the vertex.
A virtual edge is an edge without (parameter) length.

$f(G)$ is an analytic function of $G$ and
$f(G^\pm[\{\gamma\}])$ has the form
\begin{eqnarray}
&&
f(G^\pm[\{\gamma\}]):=
\nonumber\\&&{\ \ \ \ \ \ \ \ \ \ }
\sum_{n=0}^{\infty}
l_p^n\int d^3x_1\cdots d^3x_n
f_{\sigma_1\cdots\sigma_n}(x_1\cdots x_n)
G^{\sigma_1}[x_1,\{\gamma\}]\cdots G^{\sigma_n}[x_n,\{\gamma\}],
\\&&
G^\pm[x,\{\gamma\}]:=\sum_{\gamma\in\{\gamma\}} G^\pm[x,\gamma].
\end{eqnarray} 
Here $f_{\sigma_1\cdots\sigma_n}(x_1\cdots x_n)$ is a coefficient
(a function of space points)
subject to some condition.
This condition is to be determined by the 
transformation discussed in section \ref{sec:transformation}.
$\sigma_i$ ($i=1,\cdots, n$) are sumed over $+$ and $-$.
$l_p$ is the Planck length.
$\gamma\in\{\gamma\}$ means a single loop in $\{\gamma\}$.
$G^\pm[x,\gamma]$ for a single loop $\gamma$ is to be defined below.
(These states are analogues of $f(X)$ of the exact model in the 
illustration.) 
If $\Gamma$ is just a single loop $\gamma$, then 
$\langle\gamma|f(G^\pm)\rangle=f(G^\pm[\gamma])$ 
and in particular we denote
$\langle\gamma|G^\pm(x)\rangle=G^\pm[x,\gamma]$
our basis states for gravitons.
(They are analogues of $X$ of the exact model in the illustration.)

Define operators $\hat H^\pm(x)$ in the space spanned by
the family of states as follows.
\begin{eqnarray}
&&
\langle\Gamma|\hat H^\pm(x)|f(G^\pm)\rangle:=
N^{-1}(\Gamma)\sum_{\{\gamma\}\in\Gamma}
(-2)^{n(\{\gamma\})-1}(-1)^{c(\{\gamma\})} 
\times\nonumber\\&&{\ \ \ \ \ \ \ \ \ \ \ \ \ \ \ }
(-l_p^2)\int d^3yf_r(x,y)G^\pm[y,\{\gamma\}]
f(G^\pm[\{\gamma\}]).
\end{eqnarray}
Here, $f_r(x,y)$ is a function with a scale parameter $r$.
The precise definitions of this function and the scale $r$ are to be
determined by the transformation discussed in the section
\ref{sec:transformation}.
(Although these operators are not diagonal, they are analogues of
$X$ as an operator of the exact model in the illustration.)

Suppose $\Gamma$ is a spin-network and $\eta$ is a single (parameterized) 
loop without self-intersection (a special case of the spin-network).
Then $\Gamma\cup\eta$ is another spin-network consisting of $\Gamma$ and
$\eta$ if they share at most countable number of points (intersections 
between them) creating new vertices.
If they intersect with each other, then the vertex corresponding 
to one of the intersections is labeled by the singlet intertwiner (in other 
words, the trivial intertwiner or spin-0 intertwiner).
Also $\Gamma\eta-\Gamma\eta^{-1}$ is another spin-network consisting of
$\Gamma$ and $\eta$ if they share at most countable number of points
(intersections between them). 
In this case, they intersect at least once with each other.
The notation indicates that $\eta$ is attached on $\Gamma$ 
at one of the intersections in the two
different ways and the two terms are sumed with appropriate signs.
In spin-network language, the vertex corresponding to this 
intersection is labeled by one of the triplet intertwiners 
(or spin-1 intertwiners) and 
the vetices coresponding to the other intersections are labeled by
the singlet intertwiner if any.
Similarly, 
$\Gamma\eta\chi-\Gamma\eta^{-1}\chi-
\Gamma\eta\chi^{-1}+\Gamma\eta^{-1}\chi^{-1}$
is a spin-network.
$\Gamma$ shares a vertex having one of the triplet intertwiners
with $\eta$ and $\chi$ in each. 

{}From the definition of the domain states for gravitons, 
the values of the state for the spin-networks 
$\Gamma\cup\eta$, $\Gamma\eta-\Gamma\eta^{-1}$ and
$\Gamma\eta\chi-\Gamma\eta^{-1}\chi-
\Gamma\eta\chi^{-1}+\Gamma\eta^{-1}\chi^{-1}$
are determined as follows.
\begin{eqnarray}
&&
\langle\Gamma\cup\eta|f(G^\pm)\rangle=
N^{-1}(\Gamma)\sum_{\{\gamma\}\in\Gamma}
(-2)^{n(\{\gamma\}\cup\eta)-1}(-1)^{c(\{\gamma\})} 
f(G^\pm[\{\gamma\}\cup\eta]),
\\&&
\langle\Gamma_e\eta-\Gamma_e\eta^{-1}|f(G^\pm)\rangle=
N^{-1}(\Gamma)\sum_{\{\gamma\}\in\Gamma}(-2)^{n(\{\gamma\})-1}
(-1)^{c(\{\gamma\})}
\times\nonumber\\&&{\ \ \ \ \ \ \ \ \ \ }
[(2j_e)2!]^{-1}
\sum_{i\in e}\left[f(G^\pm[\{\gamma\}_i\eta])
-f(G^\pm[\{\gamma\}_i\eta^{-1}])\right],
\\&&
\langle\Gamma_{ee'}\eta\chi-\Gamma_{ee'}\eta^{-1}\chi
-\Gamma_{ee'}\eta\chi^{-1}+\Gamma_{ee'}\eta^{-1}\chi^{-1}
|f(G^\pm)\rangle=
\nonumber\\&&{\ \ \ \ \ \ \ \ \ \ }
N^{-1}(\Gamma)\sum_{\{\gamma\}\in\Gamma}(-2)^{n(\{\gamma\})-1}
(-1)^{c(\{\gamma\})}
[(2j_e)(2j_{e'})(2!)^2]^{-1}
\times\nonumber\\&&{\ \ \ \ \ \ \ \ \ \ }
\sum_{i\in e}\sum_{j\in e'}
\left[
f(G^\pm[\{\gamma\}_{ij}\eta\chi])
-f(G^\pm[\{\gamma\}_{ij}\eta^{-1}\chi])
\right.\nonumber\\&&\left.{\ \ \ \ \ \ \ \ \ \ }
-f(G^\pm[\{\gamma\}_{ij}\eta\chi^{-1}])
+f(G^\pm[\{\gamma\}_{ij}\eta^{-1}\chi^{-1}])
\right].
\end{eqnarray}
Here $e$ and $e'$ are the edges of $\Gamma$ to which $\eta$ and
$\chi$ are attached and
$i$ and $j$ are one of the segmants belonging to the edges $e$ and $e'$
respectively.
$\{\gamma\}_i\eta$ means $\eta$ is attached to the segment $i$ and 
hence one of single loops in $\{\gamma\}$ to which the segment $i$ 
belongs is modified while the other loops in $\{\gamma\}$ stay the same.

One of the basic variables of the non-perturbative theory,
called loop variables, is
\begin{eqnarray}
&&
T[\Gamma]:=N^{-1}(\Gamma)
\sum_{\{\gamma\}\in\Gamma}(-1)^{n(\{\gamma\})}
(-1)^{c(\{\gamma\})}
\prod_{\gamma\in\{\gamma\}}{\rm Tr}[\gamma].
\end{eqnarray}
${\rm Tr}[\gamma]$ is the trace of the holonomy of the Ashtekar 
connection variable along the loop $\gamma$.
Note that ${\rm Tr}[\gamma]$ goes to $2$ if $\gamma$ shrinks to a point.
If $\Gamma$ is just a single loop $\eta$, then the corresponding
variable is $T[\eta]=-{\rm Tr}[\eta]$.
Its action as an operator on the state is
\begin{eqnarray}
&&
\langle\Gamma|\hat T[\eta]:=
\langle\Gamma\cup\eta|.
\end{eqnarray}

Because of the definition of the action, 
the states satisfy the following conditions
consistently with the conventions adopted in the definition of $T[\Gamma]$.
\begin{eqnarray}
&&
\langle\Gamma\cup\eta|=\langle\Gamma^{-1}\cup\eta|=
\langle\Gamma\cup\eta^{-1}|
\\&&
\langle\Gamma\eta-\Gamma\eta^{-1}|=
-\langle\Gamma\eta^{-1}-\Gamma\eta|
\\&&
\lim_{\eta\to\cdot}\langle\Gamma\cup\eta|=-2\langle\Gamma|
\end{eqnarray}
The conventions used in this paper are similar to those adapted for
tangle theoretic recoupling theory technology \cite{tangle, deform}
and the difference does not make problems in order to understand
the main ideas of this study.
The point of the conventions is such that the loop operators
are defined locally and do not depend on how the spin-network under 
consideration is decomposed to sets of single loops.
Accordingly such non-local information of spin-networks is hidden 
in the states. 
Since our task is to define a family of states explicitly,
our states have to contain such information.

Before defining other operators and $G^\pm[x,\gamma]$, which contain
line integrals along an edge or a loop, we introduce a regularization
of parameterized loops.
The loops used in the theory are orientaion preserved reparameterization 
invariant.
Divide a loop to $N$ pieces of curves each of which has parameter length
$\rho$.
The pieces are labeled by $k=1,2,\cdots N$ 
in the order of the orientation
of the loop.
Each piece is assumed to be smooth.
If the loop has a self-intersection, then we divide
the loop such that two pieces contain the intersection point.
If the loop intersects with another loop, then we divide the loop
such that a piece of the loop shares the intersection point with
one piece of the other loop.
At the end of all the calculations, take the limits $\rho\to 0$ and
$N\to\infty$ such that $\rho N$ is finite.
We define the line integral along a loop $\gamma$ as
\begin{eqnarray}
&&
\sum_{k=1}^{N}\rho\dot\gamma^a(k)\cdots
\longrightarrow_{\rho\to 0,N\to\infty}\longrightarrow
\oint dt\dot\gamma^a(t)\cdots\equiv
\oint_\gamma dx^a\cdots
\end{eqnarray}
where $\dot\gamma^a(k)$ is the tangent vector
at the $k$-th piece of the loop $\gamma$.
We keep $\rho$ and $N$ finite untill the end of all the calculations
although $\rho$ is ``small'' and $N$ is ``large''.
The regularization for the line integral for an edge of the spin-network
is defined similarly.
In this case, the vertices at the ends of the edge are treated in the
same way as intersections of the loops are treated.

The action of another loop operator we need is defined as follows.
\begin{eqnarray}
&&
\langle\Gamma|\hat T^a[\eta](\eta(s)):=
-il_p^2\sum_{e\in\Gamma} 2j_e \sum_{t=1}^N\rho\dot\l_e^a(t)
\delta^3(\eta(s),l_e(t))
\langle\Gamma_e\eta-\Gamma_e\eta^{-1}|.
\end{eqnarray}
In terms of this operator, define another operator 
\begin{eqnarray}
&&
\hat D^{ab}(x,\delta):=-{1\over2}{1\over\pi\delta^2}\sum_{s=1}^N
\hat T^b[\eta^a_{x,\delta}](\eta(s)),
\end{eqnarray}
where $\eta^a_{x,\delta}$ is a parameterized circle 
with radius $\delta$ centered at $x$ and
normal to the $a$-direction.
This operator with the limit $\delta\to 0$ obeys the Leibnitz rule.
(This operator, or more precisely 
$\lim_{\delta\to 0}\epsilon_{abc}(\partial^b/|\partial|^2)
\hat D^{ci}(x,\delta)$,
is an analogue of $P$ of the exact model in the 
illustration.)
In the single equation, we often simplify notations for loops such as
$\eta\equiv\eta_x\equiv\eta_{x,\delta}\equiv\eta^a_{x,\delta}$
if it is clear from the contex.

We define $G^\pm[x,\gamma]$ as follows.
Here $\gamma$ and $\eta$ are single loops.
\begin{eqnarray}
&&
G^\pm[x,\gamma]:=
{1\over2}
\sum_{l}\nu_l\int d^3y\left|\sum_{k=1}^N \rho\dot\gamma^a(k)
\delta^3(\gamma(k),y)\omega_a^{\pm l}(y-x)\right|
\nonumber\\&&{\ \ \ \ \ \ \ \ \ \ }
={1\over2}\sum_{l}\nu_l\sum_{k=1}^N 
\left|\rho\dot\gamma^a(k)
\omega_a^{\pm l}(\gamma(k)-x)\right|,
\\&&
G^\pm[x,\gamma\eta]:=
{1\over2}\sum_{l}\nu_l\int d^3y
\left|\sum_{k=1}^N \rho\dot\gamma^a(k)
\delta^3(\gamma(k),y)\omega_a^{\pm l}(y-x) 
\right.\nonumber\\&&\left.
{\ \ \ \ \ \ \ \ \ \ }{\ \ \ \ \ \ \ \ \ \ }
+\sum_{k=1}^N \rho\dot\eta^b(k)\delta^3(\eta(k),y)
\omega_b^{\pm l}(y-x)\right|
\nonumber\\&&{\ \ \ \ \ \ \ \ \ \ }
=G^{\pm}[x,\gamma]+G^{\pm}[x,\eta]+
\nonumber\\&&{\ \ \ \ \ \ \ \ \ \ }{\ \ \ \ \ }
{1\over2}\rho\sum_{l}\nu_l
\left|
\dot\gamma^a(t)\omega_a^{\pm l}(\gamma(t)-x)+
\dot\eta^b(s)\omega_b^{\pm l}(\eta(s)-x)
\right|.
\end{eqnarray}
Here $\gamma\eta$ is a single loop consisting of $\gamma$ and $\eta$
with an intersection.
(Note that $\gamma\eta$ is different from $\gamma\cup\eta$.
The latter means two single loops $\gamma$ and $\eta$.)
The last term in the second equation is the contribution from the 
intersection and would vanish if one naively takes the limit 
$\rho\to 0$.
However, we do not do so until all the calculations are finished.
$t$ and $s$ are the labels of two pieces of curves containing the 
intersection.

$\omega^{\pm l}_a$ is a covector defined as follows.
\begin{eqnarray}
&&
\omega^{\pm l}_a(x):=\omega^{l}_a(\pm x)
\\&&
\sum_{l=1}^3{1\over2}\nu_l{\omega_a^l(x)\omega_b^l(x)\over
\Vert\omega^l(x)\Vert}=I_{ab}(x):=
(2\pi)^{-3}\int d^3k|k|^{-2}m_a(k)m_b(k)e^{ik\cdot x}.
\end{eqnarray}
It is known that $I_{ab}$ exists and can be computed explicitly 
\cite{vacuum}.
$\omega_a^l$ ($l=1,2,3$) are the eigen vectors of $I_{ab}$ and
can be computed for some values of $x$.
$\Vert\omega^l(x)\Vert$ is the norm of $\omega^l_a$ at $x$.
${1\over2}\nu_l\Vert\omega^l\Vert$ ($l=1,2,3$) are the eigen values of
$I_{ab}$ and 
$\nu_l$ ($l=1,2,3$) are the signs of the eigen values.

If $\Gamma$ and $\eta$ are smooth at the point they share 
and $\eta$ shrinks to the point, 
then $G^\pm[x,\gamma\eta]$ can be expanded as
\begin{eqnarray}
&&
G^\pm[x,\gamma\eta]=G^\pm[x,\gamma]+G^\pm[x,\eta]+
\nonumber\\&&{\ \ \ \ \ \ \ \ \ \ }
{1\over2}\rho\sum_{l}\nu_l{\dot\gamma(t)\cdot\omega^{\pm l}(\gamma(t)-x)
\over|\dot\gamma(t)\cdot\omega^{\pm l}(\gamma(t)-x)|}
\dot\eta(s)\cdot\omega^{\pm l}(\eta(s)-x)+\cdots.
\end{eqnarray}
Here, the dot products are simplifications of notation implying,
for example, $\dot\gamma\cdot\omega:=\dot\gamma^a\omega_a$.

$G^\pm[x,\gamma]$ shows the following properties.
Let $\gamma$ and $\gamma'$ be single loops and attach other loops
$\eta$ and $\chi$ on them. 
Assume that the four loops are smooth at the points 
any two of them share.
Then
\begin{eqnarray}
&&
G^\pm[x,\gamma\eta]-G^\pm[x,\gamma\eta^{-1}]=
\nonumber\\&&{\ \ \ \ \ \ \ \ \ \ }
\rho\sum_l\nu_l{\dot\gamma(t)\cdot\omega^{\pm l}(\gamma(t)-x)
\over|\dot\gamma(t)\cdot\omega^{\pm l}(\gamma(t)-x)|}
\dot\eta(s)\cdot\omega^{\pm l}(\eta(s)-x)+
{\cal O}(|\dot\eta|^2),
\label{eq:property-G1}
\\&&
G^\pm[x,\gamma\eta\chi]-G^\pm[x,\gamma\eta^{-1}\chi]-
G^\pm[x,\gamma\eta\chi^{-1}]+
\nonumber\\&&{\ \ \ \ \ \ \ \ \ \ }
G^\pm[x,\gamma\eta^{-1}\chi^{-1}]
=0+{\cal O}(|\dot\eta|^2)+{\cal O}(|\dot\chi|^2),
\label{eq:property-G2}
\\&& 
G^\pm[x,\gamma\eta\cup\gamma'\chi]-
G^\pm[x,\gamma\eta^{-1}\cup\gamma'\chi]-
G^\pm[x,\gamma\eta\cup\gamma'\chi^{-1}]+
\nonumber\\&&{\ \ \ \ \ \ \ \ \ \ }
G^\pm[x,\gamma\eta^{-1}\cup\gamma'\chi^{-1}]
=0+{\cal O}(|\dot\eta|^2)+{\cal O}(|\dot\chi|^2).
\label{eq:property-G3}
\end{eqnarray}
in the limit that each of the attached loops $\eta$ and $\chi$
shrinks to a point.


\section{Basis states of the semi-classical theory}
\label{sec:semi-classical}

The semi-classical theory describes large scale small fluctuations of 
gravitational field from a flat background spacetime.
The smallness of the field variables is specified by a dimensionless
possitive real parameter $\varepsilon\ll 1$.
Since the theory contains a scale parameter $l_p$, namely the Planck
length, another scale $r:=l_p/\varepsilon$ enters the theory.
This scale parameter $r$ defines the typical scale of fluctuations
of gravitational field \cite{mapM}.

The basis states of the semi-classical theory are
$F^\pm[x,\vec\alpha]$, functions of a triplet of loops
$\vec\alpha:=(\alpha^1,\alpha^2,\alpha^3)$, defined as
the Fourie transform of 
$|k|^{-2}F^\pm[k,\vec\alpha]$ by
\begin{eqnarray}
&&
{}F^\pm[x,\vec\alpha]:=
(2\pi)^{-3/2}\int d^3k|k|^{-2} F^\pm[k,\vec\alpha]e^{ik\cdot x}.
\end{eqnarray}
Here $F^\pm[k,\vec\alpha]$ are the two symmetric traceless transverse
components of $F^a[k,\alpha^i]$: namely
\begin{eqnarray}
&&
{}F^{+}[k,\vec\alpha]:=F^a[k,\alpha^i]\bar m_a(k)\bar m_i(k),
\\&&
{}F^{-}[k,\vec\alpha]:=F^a[k,\alpha^i]m_a(k)m_i(k).
\end{eqnarray}
$F^a[k,\alpha^i]$ is the Fourie transform of 
$F^a[x,\alpha^i]:=\oint_{\alpha^i}dy^a\delta^3(x-y)$
defined by
\begin{eqnarray}
&&
{}F^a[k,\alpha^i]:=
(2\pi)^{-3/2}\int d^3x F^a[x,\alpha^i]e^{-ik\cdot x}
\nonumber\\&&{\ \ \ \ \ \ \ \ \ \ }
=(2\pi)^{-3/2}\oint ds (\dot\alpha^i)^a(s)
e^{-ik\cdot\alpha^i(s)}.
\end{eqnarray}
$m_a$ amd $\bar m_a$ are polarization vectors satisfying
\begin{eqnarray}
&&
m_a(k)\bar m^a(k)=1,{\ \ }m_a(k)m^a(k)=m_a(k)k^a=0,
\\&&
\bar m_a(k)=-m_a(-k),{\ \ }\epsilon_{abc}k^am^b(k)\bar m^c(k)=-i|k|.
\end{eqnarray}
($F^\pm[x,\vec\alpha]$ are analogues of $x$ of the approximate model 
in the illustration.)

In terms of the basis states,
the domain states of the semi-classical theory are defined as follows.
\begin{eqnarray}
&&
\langle\vec\alpha|g(F^\pm)\rangle:=
g(F^\pm[\vec\alpha]):=
\nonumber\\&&{\ \ \ \ \ \ \ \ \ \ }
\sum_{n=0}^{\infty} l_p^n\int d^3x_1\cdots d^3x_n
g_{\sigma_1\cdots\sigma_n}(x_1\cdots x_n)
{}F^{\sigma_1}[x_1,\vec\alpha]\cdots F^{\sigma_n}[x_n,\vec\alpha].
\end{eqnarray} 
Here $g_{\sigma_1\cdots\sigma_n}(x_1\cdots x_n)$ is a slowly varying
function over the scale $r:=l_p/\varepsilon$ with respect to
any of its arguments.
(These states are analogues of $f(x)$ of the approximate model in
the illustration.)
The actions of the basic operators $\hat h^\pm(x)$ and
$\hat B^\pm(x)$ are defined on the states as follows.
\begin{eqnarray}
&&
\langle\vec\alpha|\hat h^\pm(x)|g(F^\pm)\rangle:=
-l_p^2\int d^3yg_r(x-y)F^\pm[y,\vec\alpha]
\langle\vec\alpha|g(F^\pm)\rangle
\\&&
\langle\vec\alpha|\hat B^\pm(x)|g(F^\pm)\rangle:=
\pm\langle\vec\alpha|{\delta g\over\delta F^\pm[x]}\rangle.
\end{eqnarray}
with $g_r(x):=(2\pi r^2)^{-3/2}\exp [-x^2/2r^2]$.
(These operators are analogues of $x$ and $p$ as operators of
the approximate model in the illustration.)

The semi-classical theory of loop quantum gravity was constructed 
\cite{linear} in terms of complex variables.
As long as the self-interaction terms in the Hamiltonian are truncated,
the linearized reality conditions are sucessfully incorporated
to determine an inner product with respect to which the annihilation and
creation operators $a$ and $a^\dagger$ satisfy the required commutation
relation $[a,a^\dagger]=1$.

However, if one tries to include the 
self-interaction terms perturbatively, then one has to take the higher
order terms in the linearized reality conditions into account;
otherwise, the definition of the Hamiltonian in terms of the linearized
operators become obscure.
This procedure spoils the required commutation relation since the
annihilation and creation operators cannot be anymore linear in the basic
linearized canonical variables in order for the inner product to
incorporate the linearized reality conditions to higher order terms.
Without well defined annihilation and creation operators, one cannot
construct the Fock space description of gravitons.

This difficulty can be easily overcomed if one uses real variables 
\cite{real} instead of complex variables.
In addition, the real variable formulation fixes chiral asymmetry of 
the complex variable formulation.
Moreover, most of the technologies developed \cite{linear} for the 
semi-classical theory are still valid and the considerations presented
in this paper can be applied to both complex and real variable 
formulations.
These facts have been recently realized \cite{loop-gravitons}.

Although it is unclear at present how the complex variable formulation can
overcome this difficulty, the complex variable formulation has some 
interesting features \cite{zapata} and should be deserved for further
investigations \cite{icn,uam}.


\section{Transformation}\label{sec:transformation}

We define an operator in terms of $\hat D^{ab}(x,\delta)$
in the non-perturbative representation state space.
\begin{eqnarray}
&&
\hat U(\vec\alpha):=
\lim_{\delta\to 0}
\exp\left[-{i\over2}\sum_i\int d^3x\epsilon_{abc}
{}F^a_\xi[x,\alpha^i]{\partial^b\over|\partial|^2}
\hat D^{ci}(x,\delta)\right],
\end{eqnarray}
with
$F^a_\xi[x,\alpha^i]:=\int d^3y g_\xi(x-y)F^a[y,\alpha^i]$.
Here, $\xi$ is a parameter such that $l_p\ll\xi\ll r:=\l_p/\varepsilon$.
(This operator is an analogue of $\exp[{i\over\hbar}x\cdot P]$
in the illustration.)

Then define a transformation $\cal N$ from the non-perturbative to
the semi-classsical state space in terms of $\hat U(\vec\alpha)$
such that
\begin{eqnarray}
&&
\langle\vec\alpha|{\cal N}=\langle\Delta|\hat U(\vec\alpha).
\end{eqnarray}
This abstract notation of transformation means that the transformation
$\cal N$ brings a state $\Psi(\Gamma)=\langle\Gamma|\Psi\rangle$
in the non-perturbative theory to a state
$\psi(\vec\alpha)=\langle\vec\alpha|\psi\rangle=
\langle\vec\alpha|{\cal N}|\Psi\rangle=
\langle\Delta|\hat U(\vec\alpha)|\Psi\rangle$
in the semi-classical theory.
Here $\Delta$ is a spin-network called ``weave'' \cite{weave}.
$\Delta$ approximates the flat space in the sense that if one computes
a geometrical quantity at large scales such as the area of a macroscopic
surface on the weave then its value coincides with the valule of the 
same quantity computed in terms of the flat space metric up to 
a small error of order $l_p/L$. $l_p$ is the Planck scale and $L$ is 
the scale of the surface under consideration.

We apply $\hat U(\vec\alpha)$ to the family of states, 
$|f(G^\pm)\rangle$, and evaluate it at the weave $\Delta$.
(The evaluation of $\hat U(\vec\alpha)$ at the weave $\Delta$ 
representing the flat space is analogous to the evaluation of
$\exp[{i\over\hbar}x\cdot P]$ at the origin $X=0$, around which
the approximate model is physically sensible as an approximation of 
the exact model in the illustratiopn.)
We perform calculations step by step below.

First, apply the operator $\hat D^{ab}(x,\delta)$ to the state 
$|f(G^{\pm})\rangle$ and evaluate it at the weave $\Delta$ 
and take the limit $\delta\to 0$.


\begin{eqnarray}
&&
\lim_{\delta\to 0}\langle\Delta|\hat D^{ab}(x,\delta)|f(G^{\pm})\rangle=
N^{-1}(\Delta)\sum_{\{\gamma\}\in\Delta}(-2)^{n(\{\gamma\})-1}
(-1)^{c(\{\gamma\})} 
\times\nonumber\\&&
\int d^3y {\delta f(G^{\pm}[\{\gamma\}])\over\delta G^\sigma[y,\{\gamma\}]}
\lim_{\delta\to 0}\left({1\over2}\hat D^{ab}(x,\delta)
G^\sigma[y,\{\gamma\}]\right).
\end{eqnarray}
Here we have used the fact that the operator $\hat D^{ab}(x,\delta)$
obeys the Leibnitz rule.
Note that the emergence of the factor of $1\over2$ in front of 
$\hat D^{ab}$ is due to the fact that the operation of the operator on
$G^\pm$ produces two terms in each action to the segment on an edge.

Next, apply the operator $\hat D^{ab}$ twice to 
the state $|f(G^{\pm})\rangle$
and evaluate it at the weave $\Delta$
and then take the limit $\delta\to 0$.


\begin{eqnarray}
&&
\lim_{\delta\to 0}\langle\Delta|
\hat D^{ab}(x,\delta)\hat D^{a'b'}(x',\delta)|f(G^{\pm})\rangle=
N^{-1}(\Delta)\sum_{\{\gamma\}\in\Delta}(-2)^{n(\{\gamma\})-1}
(-1)^{c(\{\gamma\})} 
\times\nonumber\\&&\left[
\int d^3y 
{\delta f(G^{\pm}[\{\gamma\}])\over\delta G^\sigma[y,\{\gamma\}]}
\lim_{\delta\to 0}\left({1\over4}
\hat D^{ab}(x,\delta)\hat D^{a'b'}(x',\delta)
G^\sigma[y,\{\gamma\}]\right)+
\right.\nonumber\\&&\left.
\int d^3y d^3y'
{\delta^2 f(G^{\pm}[\{\gamma\}])\over
\delta G^\sigma[y,\{\gamma\}] \delta G^{\sigma'}[y',\{\gamma\}]} 
\lim_{\delta\to 0}
\left({1\over2}\hat D^{ab}(x,\delta)G^\sigma[y,\{\gamma\}]\right)
\times\right.\nonumber\\&&\left.
\left({1\over2}\hat D^{a'b'}(x',\delta)G^{\sigma'}[y',\{\gamma\}]\right)
\right].
\end{eqnarray}
If $x'=x$, then we regularize the doubled operations at a single point
by taking the limit $x'\to x$.
This regularization means that we exclude the posibility that 
one $\hat D^{ab}$ operator acts to the small loop 
contained in another $\hat D^{a'b'}$ operator. 
In other words, we allow the operator $\hat D^{ab}$ act 
only to the spin-network under consideration.
In the same way,  
we can apply the operator $\hat D^{ab}$ more than twice to 
the state $|f(G^{\pm})\rangle$ and evaluate it at the weave $\Delta$
and then take the limit $\delta\to 0$.
They contain multiple operations of $\hat D^{ab}(x,\delta)$ to
$G^\pm[y,\{\gamma\}\in\Delta]$.

In order to proceed the calculation of the action of $\hat D^{ab}$ 
to $|f(G^\pm)\rangle$,
compute the action of $\hat D^{ab}(x,\delta)$ to 
$G^\pm[y,\{\gamma\}]$
for $\{\gamma\}\in\Delta$ and then take the limit $\delta\to 0$.


\begin{eqnarray}
&&
\lim_{\delta\to 0}
{1\over2}\hat D^{ab}(x,\delta)G^\sigma[y,\{\gamma\}\in\Delta]:=
\lim_{\delta\to 0}
\left[-{1\over4}{1\over\pi\delta^2}(-il_p^2)\right]
\times\nonumber\\&&
\sum_{e\in\Delta}
\sum_{s=1}^N\int dt \dot l_e^b(t)\delta^3(\eta^a_{x,\delta}(s),l_e(t))
\sum_{i\in e}\left(
G^\sigma[y,\{\gamma\}_i\eta_{x,\delta}]-
G^\sigma[y,\{\gamma\}_i\eta_{x,\delta}^{-1}]\right)
\nonumber\\&&
=\lim_{\delta\to 0}
\left[-{1\over4}{1\over\pi\delta^2}(-il_p^2)\right]
\sum_{e\in\Delta}
\sum_{s=1}^N\rho\int dt \dot l_e^b(t)
\delta^3(\eta^a_{x,\delta}(s),l_e(t))
\times\nonumber\\&&
2j_e\left[\sum_{l}\nu_l
{\dot l_e(t)\cdot\omega^{\sigma l}(l_e(t)-y)
\over|\dot l_e(t)\cdot\omega^{\sigma l}(l_e(t)-y)|}
\dot\eta^a_{x,\delta}(s)\cdot\omega^{\sigma l}(l_e(t)-y)
+{\cal O}(\delta^2)\right]
\nonumber\\&&
=-{1\over4}(-il_p^2)\sum_{l}\nu_l
\sum_{e\in\Delta}2j_e\int dt \dot l_e^b(t)
{\dot l_e(t)\cdot\omega^{\sigma l}(l_e(t)-y)
\over|\dot l_e(t)\cdot\omega^{\sigma l}(l_e(t)-y)|}
\omega^{\sigma l}_c(l_e(t)-y)
\times\nonumber\\&&
\lim_{\delta\to 0}{1\over\pi\delta^2}\oint ds
(\dot\eta^a_{x,\delta})^c(s)\delta^3(\eta_{x,\delta}(s),l_e(t))
\nonumber\\&&
=-{1\over4}(-il_p^2)\sum_{l}\nu_l
\sum_{e\in\Delta}2j_e\int dt \dot l_e^b(t)
{\dot l_e(t)\cdot\omega^{\sigma l}(l_e(t)-y)
\over|\dot l_e(t)\cdot\omega^{\sigma l}(l_e(t)-y)|}
\omega^{\sigma l}_c(l_e(t)-y)
\times\nonumber\\&&
\epsilon^{acf}\partial_f^x\delta^3(x,l_e(t)).
\end{eqnarray}
Here, in the first step, the definition of the action of the operator
$\hat T^a[\eta](\eta(s))$ was used and, in the second step, 
the property (\ref{eq:property-G1}) of $G^\pm$ was used on the limit of 
small $\delta$.
In the third step, the contributions from the weave and the small loop
$\eta$ were factored and,
in the last step, the limit $\delta\to 0$ was taken.
The line integral over the edges of $\Delta$ here does not have 
the regularized form since it does not make any confusion
while the line integral over the loop $\eta$ has the regularized form.
Notice the transfer of the mesure $\rho$ from the function $G^\pm$
to the integral of the operator. This process allows the non-vanishing 
contribution from the intersections of loops.

Compute the action of two $\hat D^{ab}(x,\delta)$ to $G^\pm[y,\{\gamma\}]$
and take the limit $\delta\to 0$.


\begin{eqnarray}
&&
\lim_{\delta\to 0}
{1\over4}\hat D^{ab}(x,\delta)\hat D^{a'b'}(x',\delta)
G^\sigma[y,\{\gamma\}]:=
\lim_{\delta\to 0}
\left[-{1\over4}{1\over\pi\delta^2}(-il_p^2)\right]^2
\times\nonumber\\&&
\sum_{e\in\Delta}\sum_{e'\in\Delta}
\sum_{s=1}^N\int dt \dot l_e^b(t)
\delta^3(\eta^a_{x,\delta}(s),l_e(t))
\sum_{s'=1}^N\int dt' \dot l_{e'}^{b'}(t')
\delta^3(\eta^{a'}_{x',\delta}(s'),l_{e'}(t'))
\times\nonumber\\&&
\sum_{i\in e}\sum_{j\in e'}
\left(
G^\sigma[y,\{\gamma\}_{ij}\eta_{x,\delta}\eta_{x',\delta}]-
G^\sigma[y,\{\gamma\}_{ij}\eta_{x,\delta}^{-1}\eta_{x',\delta}]-
G^\sigma[y,\{\gamma\}_{ij}\eta_{x,\delta}\eta_{x',\delta}^{-1}]+
\right.\nonumber\\&&\left.
G^\sigma[y,\{\gamma\}_{ij}\eta_{x,\delta}^{-1}\eta_{x',\delta}^{-1}]
\right)
=0.
\end{eqnarray}
The vanishing result is due to the properties 
(\ref{eq:property-G2}) and (\ref{eq:property-G3}) of $G^\pm$.
With the same reason, the action of more than two 
$\hat D^{ab}(x,\delta)$ to 
$G^\pm[y,\{\gamma\}]$ vanishes after taking the limit $\delta\to 0$.

Now, by using the action of $\hat D^{ab}[x,\delta]$ to 
$G^\pm[y,\{\gamma\}]$,
apply the exponent of the operator $\hat U(\vec\alpha)$
to $G^\pm[y,\{\gamma\}]$ for $\{\gamma\}\in\Delta$
and then take the limit $\delta\to 0$.


\begin{eqnarray}
&&
{1\over2}\lim_{\delta\to 0}
\left[-{i\over2}\sum_i\int d^3x\epsilon_{abc}
{}F^a_\xi[x,\alpha^i]{\partial^b\over|\partial|^2}
\hat D^{ci}(x,\delta)\right]
G^\sigma[y,\{\gamma\}\in\Delta]
\nonumber\\&&
=-{i\over2}\sum_i\int d^3x\epsilon_{abc}
{}F^a_\xi[x,\alpha^i]{\partial^b_x\over|\partial_x|^2}
\times\nonumber\\&&
{1\over4}il_p^2\sum_{l}\nu_l
\sum_{e\in\Delta}2j_e\int dt \dot l_e^i(t)
{\dot l_e(t)\cdot\omega^{\sigma l}(l_e(t)-y)
\over|\dot l_e(t)\cdot\omega^{\sigma l}(l_e(t)-y)|}
\omega^{\sigma l}_g(l_e(t)-y)
\epsilon^{cgf}\partial_f^x\delta^3(x,l_e(t))
\nonumber\\&&
=-{i\over2}\sum_i\int d^3x F^a_\xi[x,\alpha^i]
\times\nonumber\\&&
{1\over4}il_p^2\sum_{l}\nu_l
\sum_{e\in\Delta}2j_e\int dt \dot l_e^i(t)
{\dot l_e(t)\cdot\omega^{\sigma l}(l_e(t)-y)
\over|\dot l_e(t)\cdot\omega^{\sigma l}(l_e(t)-y)|}
\omega^{\sigma l}_a(l_e(t)-y)
\delta^3(x,l_e(t)).
\end{eqnarray}
Here, the first step was just the substitution of a previous result
and the divergencelessness of $F^a_\xi[x,\alpha^i]$ was used 
in the second step.

Now, we note that the spins of all the edges of the weave have the
same value $j_\Delta$ and use the 
following approximation formula. 
\begin{eqnarray}
&&
l_p^2\sum_{e\in\Delta}\sqrt{j_e(j_e+1)}\int dt\dot l_e^a(t)
{\dot l_e(t)\cdot v(l_e(t))\over |\dot l_e(t)\cdot v(l_e(t))|}
v_b(l_e(t))
\nonumber\\&&
{\ \ \ \ \ \ \ \ \ \ }{\ \ \ \ \ \ \ \ \ \ }{\ \ \ \ \ \ \ \ \ \ }
{\ \ \ \ \ \ \ \ \ \ }{\ \ \ \ \ \ \ }
=\int d^3x
{v^a(x)v_b(x)\over\Vert v(x)\Vert}
\left(1+{\cal O}(l_p/L)\right),
\end{eqnarray}
where $v_a$ is a covector slowly varying over the scale $L\gg l_p$
and $\Vert v\Vert$ is the norm of $v$ and
the integrals are performed over a space region of scale reasonably
larger than $l_p$. 
This approximation formula can be proved in the same manner other
approximation formulae were calculated \cite{weave,norbert}.

Here, $\omega_a^{\pm l}$ is a covector but not slowly varying over
any particular scale and hence this approximation formula cannot be
applied without further conditions.
Since, in the domain states, $G^\pm[y,\{\gamma\}]$ always appears 
together with functions $f_{\sigma_1\cdots\sigma_n}(x_1\cdots x_n)$, 
which must be a slowly varying function over the scale 
$r:=l_p/\varepsilon$ with respect to any of its arguments
if they are successfully transmitted to the semi-classical theory,
we restrict ourselves to the case that $\omega_a^{\pm l}(l_e(t)-y)$
is smeared out over the scale $r$ with respect to $y$.
The integrals are performed over regions of scale $\xi\gg l_p$.
Then


\begin{eqnarray}
&&
{1\over2}\lim_{\delta\to 0}
\left[-{i\over2}\sum_i\int d^3x\epsilon_{abc}
{}F^a_\xi[x,\alpha^i]{\partial^b\over|\partial|^2}
\hat D^{ci}(x,\delta)\right]
G^\sigma[y,\{\gamma\}\in\Delta]
\nonumber\\&&
=-{i\over2}\sum_i\int d^3x F^a_\xi[x,\alpha^i]
{2j_\Delta\over\sqrt{j_\Delta(j_\Delta+1)}}
{i\over2}\delta^{id}
\times\nonumber\\&&
\int d^3z \sum_{l}{1\over2}\nu_l
{\omega^{\sigma l}_d(z-y)\omega^{\sigma l}_a(z-y)
\over\Vert\omega^{\sigma l}(z-y)\Vert}
\left(1+{\cal O}(\varepsilon)\right)
\delta^3(x,z)
\nonumber\\&&
={1\over4}\sum_i\int d^3xF^a_\xi[x,\alpha^i]
{2j_\Delta\over\sqrt{j_\Delta(j_\Delta+1)}}
\delta^{id}
\sum_{l}{1\over2}\nu_l
{\omega^{\sigma l}_d(x-y)\omega^{\sigma l}_a(x-y)
\over\Vert\omega^{\sigma l}(x-y)\Vert}
\left(1+{\cal O}(\varepsilon)\right)
\nonumber\\&&
={1\over4}
{2j_\Delta\over\sqrt{j_\Delta(j_\Delta+1)}}
\sum_i\int d^3xF^a_\xi[x,\alpha^i]
\left(1+{\cal O}(\varepsilon)\right)
\times\nonumber\\&&
(2\pi)^{-3}\int d^3k|k|^{-2}m^i(k)m_a(k)e^{\sigma ik\cdot(x-y)}
\nonumber\\&&
={1\over4}
{2j_\Delta\over\sqrt{j_\Delta(j_\Delta+1)}}
\sum_i 
(2\pi)^{-3/2}\int d^3k|k|^{-2}m^i(k)m_a(k)
{}F^a_\xi[-\sigma k,\alpha^i]
e^{-\sigma ik\cdot y}
\left(1+{\cal O}(\varepsilon)\right)
\nonumber\\&&
={1\over4}
{2j_\Delta\over\sqrt{j_\Delta(j_\Delta+1)}}
(2\pi)^{-3/2}\int d^3k|k|^{-2}F^\sigma_\xi[k,\vec\alpha]e^{ik\cdot y}
\left(1+{\cal O}(\varepsilon)\right)
\nonumber\\&&
={1\over4}{2j_\Delta\over\sqrt{j_\Delta(j_\Delta+1)}}
{}F^\sigma_\xi[y,\vec\alpha]
\left(1+{\cal O}(\varepsilon)\right).
\end{eqnarray}
Here, in the first step, the approximation formula was used to
replace the sum of the line integrals along the edges of the weave
by an integral over space $z$ and the integration was done in 
the second step.
In the third step, the definition of $\omega^{\pm l}_a$ 
in terms of the polarization vectors was used
and, subsequently, Fourie transformations have been repeated to find
the final line.

Before evaluating the transformation,
apply the exponent of the operator $\hat U[\vec\alpha]$
to $|f(G^\pm)\rangle$ and evaluate it at the weave $\Delta$
and take the limit $\delta\to 0$.


\begin{eqnarray}
&&
\lim_{\delta\to 0}\langle\Delta|
\left[-{i\over2}\sum_i\int d^3x\epsilon_{abc}
{}F^a_\xi[x,\alpha^i]{\partial^b\over|\partial|^2}
\hat D^{ci}(x,\delta)\right]
|f(G^\pm)\rangle=
\nonumber\\&&
N^{-1}(\Delta)\sum_{\{\gamma\}\in\Delta}(-2)^{n(\{\gamma\})-1}
(-1)^{c(\{\gamma\})} 
\int d^3y {\delta f(G^{\pm}[\{\gamma\}])\over
\delta G^\sigma[y,\{\gamma\}]}
\lambda_1(\Delta,\omega^\pm)
{}F^\sigma_\xi[y,\vec\alpha],
\end{eqnarray}
where
\begin{eqnarray}
\lambda_1(\Delta,\omega^\pm):=
{1\over4}{2j_\Delta\over\sqrt{j_\Delta(j_\Delta+1)}}
\left(1+{\cal O}(\varepsilon)\right),
\end{eqnarray}
which is a number of order $1$.
$\cal O(\varepsilon)$ depends on $\omega^{\pm l}_a$ and $\Delta$.

In the same way,
apply the exponent of the operator $\hat U[\vec\alpha]$ twice
to $|f(G^\pm)\rangle$ and evaluate it at the weave $\Delta$
and take the limit $\delta\to 0$.


\begin{eqnarray}
&&
\lim_{\delta\to 0}\langle\Delta|
{1\over2}\left[-{i\over2}\sum_i\int d^3x\epsilon_{abc}
{}F^a_\xi[x,\alpha^i]{\partial^b\over|\partial|^2}
\hat D^{ci}(x,\delta)\right]^2
|f(G^\pm)\rangle=
\nonumber\\&&
N^{-1}(\Delta)\sum_{\{\gamma\}\in\Delta}(-2)^{n(\{\gamma\})-1}
(-1)^{c(\{\gamma\})} 
\times\nonumber\\&&
{1\over2}\int d^3y d^3y' 
{\delta^2 f(G^{\pm}[\{\gamma\}])\over
\delta G^\sigma[y,\{\gamma\}]\delta G^{\sigma'}[y',\{\gamma\}]}
\lambda_1 F^\sigma_\xi[y,\vec\alpha]
\lambda_1 F^{\sigma'}_\xi[y',\vec\alpha].
\end{eqnarray}

In general, the $n$-th power of
the exponent of the operator $\hat U(\vec\alpha)$ applying
to $|f(G^\pm)\rangle$ at the weave $\Delta$ 
with the limit $\delta\to 0$
can be computed as follows.


\begin{eqnarray}
&&
\lim_{\delta\to 0}\langle\Delta|
\left[-{i\over2}\sum_i\int d^3x\epsilon_{abc}
{}F^a_\xi[x,\alpha^i]{\partial^b\over|\partial|^2}
\hat D^{ci}(x,\delta)\right]^n
|f(G^\pm)\rangle=
\nonumber\\&&
N^{-1}(\Delta)\sum_{\{\gamma\}\in\Delta}(-2)^{n(\{\gamma\})-1}
(-1)^{c(\{\gamma\})} 
\times\nonumber\\&&
\int d^3y_1\cdots d^3y_n 
{\delta^n f(G^{\pm}[\{\gamma\}])\over
\delta G^{\sigma_1}[y_1,\{\gamma\}]\cdots
\delta G^{\sigma_n}[y_n,\{\gamma\}]}
\lambda_1 F^{\sigma_1}_\xi[y_1,\vec\alpha]\cdots
\lambda_1 F^{\sigma_n}_\xi[y_n,\vec\alpha].
\end{eqnarray}

{}From these results, it is easy to evaluate the transformation
to the state $|f(G^{\pm})\rangle$ at the weave $\Delta$.


\begin{eqnarray}
&&
\langle\Delta|
\hat U(\vec\alpha)|f(G^\pm)\rangle=
N^{-1}(\Delta)\sum_{\{\gamma\}\in\Delta}(-2)^{n(\{\gamma\})-1}
(-1)^{c(\{\gamma\})} 
\times\nonumber\\&&
\sum_{n=0}^\infty{1\over n!}\int d^3y_1\cdots d^3y_n 
{\delta^n f(G^{\pm}[\{\gamma\}])\over
\delta G^{\sigma_1}[y_1,\{\gamma\}]\cdots
\delta G^{\sigma_n}[y_n,\{\gamma\}]}
\lambda_1 F^{\sigma_1}_\xi[y_1,\vec\alpha]\cdots
\lambda_1 F^{\sigma_n}_\xi[y_n,\vec\alpha]
\nonumber\\&&
=C(\Delta) f(\lambda_0(\Delta,\omega^\pm)+
\lambda_1(\Delta,\omega^\pm) F^\pm_\xi[\vec\alpha]).
\end{eqnarray}
Here $C(\Delta)$ is a multiplicative constant depending on the weave
defined by
\begin{eqnarray}
&&
C(\Delta):=
N^{-1}(\Delta)\sum_{\{\gamma\}\in\Delta}(-2)^{n(\{\gamma\})-1}
(-1)^{c(\{\gamma\})}, 
\end{eqnarray}
and
$f(\lambda_0+\lambda_1 F^\pm_\xi[\vec\alpha])$ is 
$f(G^\pm[\{\gamma\}])$ with $G^\pm[x,\{\gamma\}]$ replaced by
$\lambda_0+\lambda_1 F^\pm_\xi[x,\vec\alpha]$. 
$\lambda_0$ is defined for $\{\gamma\}\in\Delta$ by
\begin{eqnarray}
&&
\lambda_0(\Delta,\omega^\pm):=
G^\pm[x,\{\gamma\}\in\Delta]
\longrightarrow_{\rho\to 0,N\to\infty}\longrightarrow
\nonumber\\&&
\sum_{e\in\Delta}2j_e\sum_{l}{1\over2}\nu_l\int dt
|\dot l_e^a(t)\omega^{\pm l}_a(l_e(t)-x)|
\nonumber\\&&
={2j_\Delta\over\sqrt{j_\Delta(j_\Delta+1)}}
\sum_{l}{1\over2}\nu_l
\sum_{e\in\Delta}\sqrt{j_e(j_e+1)}
\int dt|\dot l_e^a(t)\omega^{\pm l}_a(l_e(t)-x)|
\nonumber\\&&
=l_p^{-2}{2j_\Delta\over\sqrt{j_\Delta(j_\Delta+1)}}
\sum_{l}{1\over2}\nu_l
\int dz\Vert\omega^{\pm l}(z-x)\Vert
\left(1+{\cal O}(\varepsilon)\right)
\nonumber\\&&
=l_p^{-2}{2j_\Delta\over\sqrt{j_\Delta(j_\Delta+1)}}
\sum_{l}{1\over2}\nu_l
\int dz\Vert\omega^{\pm l}(z-x)\Vert
{\cal O}(\varepsilon),
\end{eqnarray}
where the leading term has vanished because of
$\sum_l{1\over2}\nu_l\Vert\omega^{\pm l}\Vert=0$ at each space point
while the other term has survived since the error
${\cal O}(\varepsilon)$ depends on space point implicitly through
$\omega^{\pm l}_a$ and $\Delta$.
In the computation of $\lambda_0$, we have used another approximation 
formula as follows.
\begin{eqnarray}
&&
l_p^2\sum_{e\in\Delta}\sqrt{j_e(j_e+1)}\int dt
|\dot l_e^a(t)v_a(l_e(t))|=\int d^3x\Vert v(x)\Vert
\left(1+{\cal O}(l_p/L)\right).
\end{eqnarray}
Again, as in the other approximation formula,
$v_a$ is a covector slowly varying over the scale $L\gg l_p$
and $\Vert v\Vert$ is the norm of $v$. 
The integrals are performed over a space region of scale reasonably
larger than $l_p$.
This approximation formula is already known \cite{weave,norbert},
or more precisely, the weave is defined such a way that 
this approximation formula holds.

The transformation $\cal N$ transforms an analytic function of $G^\pm$
to the same analytic function with $G^\pm$ replaced by 
$\lambda_0+\lambda_1 F^\pm_\xi$
provided that the coefficient functions such as 
$f_{\sigma_1\cdots\sigma_n}(x_n\cdots x_n)$
are slowly varying functions over the scale $r$ with respect to
any of their arguments.
In particular, $G^\pm$ itself is transformed to
$\lambda_0+\lambda_1 F^\pm_\xi$, if it is smeared against a slowly
varying function over the scale $r$, as follows.
\begin{eqnarray}
&&
\lambda_0(\Delta,\omega)+\lambda_1(\Delta,\omega) F^\pm_\xi[x,\vec\alpha]
=\langle\Delta|\hat U(\vec\alpha)|G^\pm(x)\rangle.
\end{eqnarray}
Here $\lambda_1$ and $\lambda_0$ are of order $1$ and $\varepsilon$
respectively and
depend on the weave and $\omega^{\pm l}$. 
These terms are 
Planck scale corrections coming from the difference between the discrete 
structure of the weave and the flat space structure assumed at all the 
scales in the semi-classical theory.

Now, what is of interest is that the Plank scale correction terms appear
always with $F^\pm$. 
One is multiplicative to and the other additive to $F^\pm$.
We notice that in the semi-classical theory the Planck length constant
enters the theory always multiplicative to $F^\pm$ and 
a constant field additive to $F^\pm$ defines a unitarily equivalent
theory.
Therefore, the multiplicative correction can be absorbed to the 
original Planck length constant $l_p$ to redefine 
a ``coarse-grained'' constant $\bar l_p$.
Then the additive correction term can be eliminated by a unitary
transformation of the variables of the semi-classical theory.

After all, the transformation $\cal N$ transforms an analytic function 
of $G^\pm$ to the same analytic function with $G^\pm$ replaced 
by $F^\pm_\xi$ and with $l_p$ replaced by a ``coarse-grained'' constant
$\bar l_p$ provided that the coefficient functions contained are
slowly varying functions over the scale $r$.
Here the parameter $\xi$ is so small compared to the scale of graviton
fluctuations $r$, we disregard the difference between
$F^\pm$ and $F^\pm_\xi$on the physical ground and simply consider
$F^\pm_\xi$ as a regularization form of $F^\pm$.
Accordingly, the multiplication of and the derivative with respect to 
$G^\pm$ respectively correspond through the transformation
to the multiplication of 
and the derivative with respect to $F^\pm_\xi$.
Since the multiplication of and the derivative with respect to 
$F^\pm$ are the basic operations of the semi-classical theory,
the operators $\hat H^\pm$ and $\hat D^{ab}$ constructed in section 
\ref{sec:non-perturbative} provide the basic operations on the
domain states for gravitons.
$\hat H^\pm$ and suitable components of $\hat D^{ab}$ define respectively
the multiplication of and the derivative with repsect to $G^\pm$
in the non-perturbative theory in the same sense that $\hat h^\pm$
and $\hat B^\pm$ define respectively the multiplication of and the
derivative with respect to $F^\pm$ in the semi-classical theory
provided that $f_r(x,y)$ in $\hat H^\pm$ is identified to $g_r(x-y)$
in $\hat h^\pm$.


\section{Comparison with the previous work}\label{sec:comparison}

In this section, we compare the present work with the previous work 
\cite{mapM,vacuum}.
In particular, we realize how the transformation $\cal N$ constructed in
the present work improves the interpretation of the graviton states
in the non-pertubative theory.
The interpretation was partially made in the previous work.

\subsection{The previous work: the transformation $\cal M$}

In the previous work,
a transformation denoted by $\cal M$ was constructed.
It transforms a family of states, the prototype of the family of states
presented in the present work,
in the non-pertubative loop representation
space to states in the semi-classical loop representation space.
However, $\cal M$ contains an error term due to its definition and 
its use is limited to transformations of states and operators 
of the first order magnitude in the fluctuations of gravitatinal field
in terms of suitable norm. 
{}From this limitation the functions $G^\pm[x,\gamma]$ and some operator 
$\hat A$ were found to be related respectively to the basis 
states $F^\pm[x,\vec\alpha]$ 
(the Fourie transform of the symmetric traceless
transverse components of $F^a[k,\alpha^i]$) and the annihilation operator 
$\hat a$ (a linear combination of the basic canonical variables) 
of the semi-classical theory, that is
\begin{eqnarray}
&&
F^\pm={\cal M}G^\pm+{\cal O}(\varepsilon^2)
\\&&
\hat a{\cal M}={\cal M}\hat A+{\cal O}(\varepsilon^2).
\end{eqnarray}
Here $\varepsilon$ is the order of the fluctuations of gravitational field.

By interpreting $G^\pm$ and $\hat A$ as basis states and annihilation 
operator in the non-perturbative loop representation space respectively,
the graviton states were constructed INDIRECTLY.
For example, The ground state was found by requiring that it is annihilated
by the annihilation operator $\hat A$ up to an error of order $\varepsilon^2$
and it is a function of $G^\pm$.

\subsection{The present work: the transformation $\cal N$}

In the present work, a transformaton denoted by $\cal N$ was constructed.
It transforms a family of states containing a function of $G^\pm$
in the non-perturbative loop representaion space
to the state with the form of the same function of $F^\pm_\xi$ in the 
semi-classical loop represenation space WITHOUT an error term, that is
\begin{eqnarray}
&&
f(\lambda_1 F^\pm_\xi +\lambda_0)={\cal N}f(G^\pm).
\end{eqnarray}
The constants $\lambda_0$ and $\lambda_1$ have emerged as  ``deviations''
of the weave structure from flat space.
By redefining the Planck length constant in the semi-classical theory
and performing a unitary transformation of the canonical variables
to eliminate $\lambda_1$ and $\lambda_0$ respectively,
we found
\begin{eqnarray}
&&
f(F^\pm_\xi)={\cal N}f(G^\pm).
\end{eqnarray}
{}From this we can immediately find the graviton states in the 
non-perturbative theroy as before 
but DIRECTLY without solving any requirement
since they are just special cases of functions of $G^\pm$.
 
An important difference is the following.
The transformation $\cal N$ does not contain an error term.
Therefore, the definition of the graviton states in the non-perturbative 
theroy is ``exact'' although they are approximate states in the sense
that they have nothing to do with the exact constraints
of the non-perturbative theory.
The transformation $\cal M$ allows the interpretation that these
states represent free gravitons up to an error of order
$\varepsilon^2$ because of the presence of an error term in it.
This error term spoils the interpretation when the self-interaction of
gravitons is introduced possibly perturbatively or non-perturbatively.
The transformation $\cal N$ keeps the interpretation valid even after
introducing the self-interaction because of the absence of error term.

\section{Conclusions}\label{sec:conclusions}

We have constructed, in the non-perturbative loop representation space,
what we call basis states for gravitons, denoted by $G^\pm$, 
in terms of which a family of states are defined.
We have showed that there exists a transformation which transforms
the family of states to the domain states of the semi-classical
loop representation space.
Given a state made of an analytic function of $G^\pm$ 
with a parametr $l_p$ (the Planck length constant), 
denoted by $f(G^\pm)$, the transformation transforms it to the same
analytic function with $G^\pm$ and $l_p$ replaced by $F^\pm_\xi$
(the basis states of the semi-classical loop representation space)
and $\bar l_p$ (a ``coars-grained'' constant) respectively up to
an overall multiplicative constant.

Therefore, all the domain states of the
semi-classical loop representation space can be recovered, 
through the transformation $\cal N$, from
the family of states in the non-perturbative loop representation space.
In particular, the states corresponding to the graviton states
of the semi-classical theory can be found easily since they are
special cases of the family of states.
They are the graviton states in the non-perturbative theory.
This relation of the two theories is ``exact'' in the sense that
the inclusion of self-interaction of gravitons does not make the
interpretation of these graviton states obscure.

Although to prove a mathematical exactness such as isomorphism of
some sectors of the two theories is not clear,
the existence of this ``exact'' relation of the two theories supports
the potential ability of the non-perturbative loop representation 
quantum gravity
to address the physics of gravitons, namely quanta for large scale small
fluctuations of gravitational field from the flat background spacetime.  

\section*{Acknowlegements}

I would like to thank the members of
the Department of Gravitation and Field Theory
in Instituto de Ciencias Nucleares,
Universidad Nacional Autonoma de Mexico, Mexico
and of
the Gravity group of Department of Physics in
Universidad Autonoma Metropolitana, Iztapalapa, Mexico
for hospitality.


\end{document}